\documentclass[12pt]{article}
\usepackage[left=2.0cm,top=2.0cm,right=2.0cm,bottom=2.0cm]{geometry}
\usepackage{amsmath, amssymb}
\usepackage[latin1]{inputenc}
\usepackage{mathrsfs}
\usepackage{cite}
\usepackage{dsfont}
\usepackage{graphicx}

\begin{document}
\sloppy
\date{}
\title{Exact Solution of the $(2+1)$-Dimensional Dunkl-Klein-Gordon Oscillator with a Uniform Magnetic Field in Non-Commutative Phase Space}

\author{R. D. Mota$^{a}$, and  D. Ojeda-Guill\'en$^{b}$\footnote{{\it E-mail address:} dojedag@ipn.mx}} \maketitle

\begin{minipage}{0.9\textwidth}

\small $^{a}$ Escuela Superior de Ingenier{\'i}a Mec\'anica y El\'ectrica, Unidad Culhuac\'an,
Instituto Polit\'ecnico Nacional, Av. Santa Ana No. 1000, Col. San
Francisco Culhuac\'an, Del. Coyoac\'an, C.P. 04430, Ciudad de M\'exico, Mexico.

\small $^{b}$ Escuela Superior de C\'omputo, Instituto Polit\'ecnico Nacional,
Av. Juan de Dios B\'atiz esq. Av. Miguel Oth\'on de Mendiz\'abal, Col. Lindavista,
Del. Gustavo A. Madero, C.P. 07738, Ciudad de M\'exico, Mexico.
\end{minipage}

\begin{abstract}
We solve the Klein-Gordon oscillator in a non-commutative phase space, generalized using Dunkl derivatives, both with and without interaction with a uniform magnetic field. For each of these cases, we begin by introducing the Dunkl derivative into the standard commutative phase space to evaluate the reflection parity symmetries. Then, using the Seiberg-Witten map, we extend the generalized system to a non-commutative phase space. In both cases, we determine the eigenfunctions and the energy spectrum exactly (which correspond to generalized Landau levels when a magnetic field is present). Furthermore, we compute the non-relativistic limit of the system and graphically analyze the behavior of the energy spectra. Our results demonstrate that the combination of the Dunkl deformation parameters, the non-commutative phase space geometry, and the magnetic field produces a parity-dependent modification of the energy levels. Finally, we verify that when the external magnetic field or the non-commutative parameters vanish, our exact solutions correctly reduce to the commutative Dunkl models established in previous works.

\vspace{0.5cm}
\end{abstract}

\section{Introduction}

A problem of fundamental interest in relativistic quantum mechanics is a charged particle coupled to an external uniform magnetic field. Unlike the Dirac oscillator, which describes spin-$1/2$ fermions, the Klein-Gordon oscillator is the canonical model for describing the dynamics of spin-$0$ scalar bosons, such as pions or mesonic condensates \cite{Greiner2000}. In the standard commutative space, this dynamics is formulated through the Klein-Gordon oscillator and its coupling to magnetic fields, whose exact solutions yield the well-known oscillator and scalar Landau energy levels \cite{Bruce1993,Dvoeglazov1994,Rao2008,Boumali2011,Bakke2015}. Because scalar bosons do not possess the inherent spin-magnetic coupling (Zeeman splitting) present in fermions, the Klein-Gordon oscillator provides an ideal framework to isolate and study purely orbital dynamical effects.

The mathematical generalization of quantum observables through reflection symmetries began with Wigner's formulation of modified commutation relations \cite{Wigner1950}. This concept was subsequently materialized by Yang in the context of the harmonic oscillator \cite{Yang1951} and later applied to the exact solvability of Calogero models \cite{Brzezinski1993}. A rigorous mathematical framework for these symmetries was established by Dunkl \cite{Dunkl1989}, whose differential-difference operators successfully eliminate the inconsistencies encountered by previous algebraic deformations when acting on constant functions \cite{Watanabe1989}.

Since their introduction, Dunkl operators have become a fundamental tool in the analysis of central potentials and the Laplace operator \cite{Dunkl2001}. In recent years, this formalism has experienced a significant increase in applications in quantum mechanics, starting with the exact resolution and superintegrability analysis of non-relativistic systems such as the isotropic oscillator and the Coulomb potential \cite{Genest2013,Genest2014a,Genest2014b,Genest2015,Nos2017,Nos2018,Ghazouani2019a,Ghazouani2019b}. The natural progression of this framework led to its successful implementation in relativistic wave equations, including the Klein-Gordon and Dirac systems under various configurations \cite{Nos2019a,Nos2020,Nos2021,Hassanabadi2022,Hamil2022,Sedaghatnia2022}. Currently, the Dunkl deformation is being extensively explored across diverse branches of theoretical physics, ranging from few-body dynamics to thermodynamic properties and alternative relativistic wave equations \cite{Chung2023a,Chung2023b,Junker2023,Bouguerne,Schulze2023,Schulze2024a,Schulze2024b,Schulze2024c,Junker2024,
Schulze2025,Hamil2025a,Nos2024a,Nos2024b,Khantoul2026,Medani2026,Hassanabadi2024,Benzair2025a,Benchikha2025,
Sedaghatnia2025,Benzair2025b,Lut2025,Hocine2025,Benzair2026,Salazar2026a,Arabsaghari2024,Chung2023c,Chung2021,
Sedaghatnia2023,Benzair2024}.

Parallel to these developments, the formulation of quantum mechanics in non-commutative phase spaces has received considerable attention as a way to explore fundamental length scales. Analytical solutions for various relativistic and non-relativistic oscillators in these deformed spaces reveal how spatial and momentum non-commutativity deeply modify the corresponding energy spectra \cite{Cai2010,Cai2011,Jian-Hua2008,Mandal2012,Panella2014,Panahi2017}. Moreover, recent literature has extensively investigated relativistic quantum systems, namely spin-zero bosons and generalized oscillators under various algebraic and topological backgrounds, including non-commutative geometries, Lorentz symmetry violations, and topological defects such as global monopoles and cosmic dislocations \cite{Hassanabadi2018,Zare2020,Montigny2021a,Montigny2021b,Zare2022a,Zare2022b,Zare2022c,Hassanabadi2023}.

Despite these significant advancements, the study of relativistic scalar bosons interacting simultaneously with an external magnetic field, non-commutative phase space deformations, and Dunkl reflection symmetries remains completely unexplored. To bridge this gap, in the present work we study the $(2+1)$-dimensional Dunkl-Klein-Gordon oscillator coupled to a uniform external magnetic field in a non-commutative phase space. The novelty of this work lies in the rigorous sequential integration of these formalisms. By evaluating the commutators using the Dunkl algebra prior to the phase space deformation, we determine how the simultaneous interplay between the non-commutative parameters ($\theta, \eta$), the external magnetic field ($B$), and the parity-dependent Dunkl parameters ($\mu_x, \mu_y$) modifies the scalar Landau energy levels. Consequently, our work goes a step further than previous studies that investigated either the non-commutative relativistic oscillator with standard derivatives \cite{Cai2010,Cai2011,Jian-Hua2008}, or the Dunkl-Klein-Gordon oscillator in commutative space \cite{Nos2021,Nos2021b}.

This paper is organized as follows. In Section 2, we start with the standard Klein-Gordon oscillator and apply the Dunkl derivative in a commutative space. Then, we use the Seiberg-Witten map to extend the system to a non-commutative phase space. For this case without a magnetic field, we find the exact solutions and the energy spectrum, and we show how our results reduce to the standard limits. In Section 3, we add a uniform magnetic field to the system. We calculate the exact wave functions and the generalized Landau levels for different parity states. Also, we study the special limiting and non-relativistic cases, and present a graphical analysis of the energy spectra. Finally, Section 4 presents our conclusions.

\section{The Dunkl-Klein-Gordon oscillator in non-commutative phase space}

In this section, we discuss the Klein-Gordon oscillator, generalized first by including the Dunkl derivative and then extended to a non-commutative phase space. Since we do not consider a uniform magnetic field in this stage, we refer to this system as the standard Dunkl-Klein-Gordon oscillator in a non-commutative phase space.

\subsection{Differential equation of the Klein-Gordon oscillator}

In the commutative space, the standard commutation relations $[x_i, x_j] = 0$, $[p_i, p_j] = 0$, and $[x_i, p_j] = i\hbar \delta_{ij}$ are satisfied. Under these conditions, the Klein-Gordon oscillator is described by the equation \cite{Bruce1993,Dvoeglazov1994,Rao2008,Boumali2011,Bakke2015}
\begin{equation}
c^2(\mathbf{p} + i m \omega \boldsymbol{\rho}) \cdot (\mathbf{p} - i m \omega \boldsymbol{\rho}) \psi(\boldsymbol{\rho}) = \left( E^2 - m^2 c^4\right) \psi(\boldsymbol{\rho}).\label{kgo}
\end{equation}
Expanding the left-hand product, we can write this equation in the form
\begin{equation}
\left[ p^2 + m^2 \omega^2 \rho^2 - i m \omega \left( (p_x x - x p_x) + (p_y y - y p_y) \right) \right] \psi(x,y) = \left( \frac{E^2 - m^2 c^4}{c^2} \right) \psi(x,y).\label{nckg}
\end{equation}

\subsection{The Dunkl-Klein-Gordon differential equation in commutative space}

To construct the generalized Dunkl-Klein-Gordon oscillator, our first step is to introduce the Dunkl formalism into the standard commutative phase space. We replace the standard partial derivative with the Dunkl derivative, which is defined by the expression \cite{Dunkl1989,Genest2013}
\begin{equation}
D_i = \partial_i + \frac{\mu_i}{x_i}(1 - R_i), \hspace{5ex}i=1,2.
\end{equation}
Here, $\mu_i > -1/2$ is a real parameter and $R_i$ is the reflection operator with respect to the $x_i$-axis (where $x_1\equiv x$ and $x_2\equiv y$), such that $R_x f(x,y) = f(-x,y)$. The Dunkl momentum is defined as $p_i^D=-i\hbar D_i$. Therefore, the Dunkl angular momentum and the Dunkl Laplacian take the respective forms \cite{Genest2013,Genest2014a}
\begin{eqnarray}
&&L_z^D = x p_y^D - y p_x^D= -i\hbar \left(xD_y-yD_x\right)\equiv -\hbar \mathcal{J},\label{ang}\\
&&(p^D)^2 = (p_x^D)^2 + (p_y^D)^2=-\hbar^2 \left(D_x^2+D_y^2\right)\equiv -\hbar^2\nabla_D^2.\label{lapla}
\end{eqnarray}
Additionally, the fundamental commutation relation of the Dunkl algebra follows immediately and is written as
\begin{equation}
 [D_i, x_i] = 1 + 2\mu_i R_i, \hspace{5ex} i=1,2. \label{com}
\end{equation}

By substituting the standard momentum $p$ with the Dunkl momentum $p^D$ in Eq. (\ref{nckg}), the cross-terms are evaluated using the Dunkl commutation relations. Specifically, $p_x^D x - x p_x^D = -i\hbar[D_x, x] = -i\hbar(1 + 2\mu_x R_x)$. Consequently, the interaction terms simplify exactly to
\begin{equation}
-i m \omega \left( (p_x^D x - x p_x^D) + (p_y^D y - y p_y^D) \right) = -2 m \omega \hbar (1 + \mu_x R_x + \mu_y R_y).
\end{equation}
Thus, the differential equation for the Dunkl-Klein-Gordon oscillator in a commutative space takes the form
\begin{equation}
\left[ (p^D)^2 + m^2 \omega^2 \rho^2 - 2 m \omega \hbar (1 + \mu_x R_x + \mu_y R_y) \right] \psi(x,y) = \left( \frac{E^2 - m^2 c^4}{c^2} \right) \psi(x,y). \label{dkg_comm}
\end{equation}

\subsection{Non-commutative Dunkl-Klein-Gordon oscillator via Seiberg-Witten map}

We now extend the generalized Dunkl-Klein-Gordon oscillator to a non-commutative phase space. In a standard non-commutative phase space, the continuous dynamic variables satisfy the following commutation relations
\begin{equation}
\left[\hat{x},\hat{y}\right]=i\theta,\hspace{5ex}\left[\hat{p}_x,\hat{p}_y\right]=i\eta,\hspace{5ex}\theta, \eta \in \mathbb{R}.
\end{equation}
Since the reflection operators $R_x$ and $R_y$ represent discrete parity symmetries and are not continuous dynamic variables of the phase space, they remain invariant under this deformation. Consequently, we propose the differential equation for the oscillator in the non-commutative extension of Eq. (\ref{dkg_comm}) becomes
\begin{equation}
\left[ (\hat{p}^D)^2 + m^2 \omega^2 (\hat{\rho}^{D})^2 - 2 m \omega \hbar (1 + \mu_x R_x + \mu_y R_y) \right]\star \psi(x,y) = \left( \frac{E^2 - m^2 c^4}{c^2} \right) \psi(x,y),
\end{equation}
where the hats emphasize that the continuous variables are non-commutative, with $(\hat{p}^D)^2 = (\hat{p}_x^D)^2 + (\hat{p}_y^D)^2$ and $(\hat{\rho}^{D})^2 = (\hat{x}^{D})^2 + (\hat{y}^{D})^2$.

To transform this non-commutative formulation back into a more accessible commutative framework, we employ a phase space transformation. In this work, we use the Seiberg-Witten map to express the non-commutative coordinates and momenta in terms of their commutative Dunkl counterparts
\begin{eqnarray}
&&\hat{x}^{D} = x - \frac{\theta}{2\hbar} p_y^D, \hspace{8ex}\hat{y}^{D} = y + \frac{\theta}{2\hbar} p_x^D \label{wm1}\\
&&\hat{p}_x^D = p_x^D + \frac{\eta}{2\hbar} y, \hspace{7ex}\hat{p}_y^D = p_y^D - \frac{\eta}{2\hbar} x.\label{wm2}
\end{eqnarray}
Applying these transformations, the squared operators expand as
\begin{eqnarray}
&&(\hat{p}^D)^2 = (p^D)^2 + \frac{\eta^2}{4\hbar^2}\rho^2 - \frac{\eta}{\hbar}L_z^D, \\
&&(\hat{\rho}^{D})^2 = \rho^2 + \frac{\theta^2}{4\hbar^2}(p^D)^2 - \frac{\theta}{\hbar}L_z^D.
\end{eqnarray}
Substituting these expansions into the non-commutative equation, grouping like terms, and using the definitions $(p^D)^2 = -\hbar^2 \nabla_D^2$ and $L_z^D = -\hbar \mathcal{J}$, we obtain the following equation for the non-commutative Dunkl-Klein-Gordon oscillator
\begin{multline}
H_O\psi(x,y)\equiv\Bigg[ -\hbar^2\left( 1 + m^2 \omega^2 \frac{\theta^2}{4\hbar^2} \right) \nabla_D^2 + \left( m^2 \omega^2 + \frac{\eta^2}{4\hbar^2} \right) \rho^2 +\left( \eta + m^2 \omega^2 \theta \right) \mathcal{J}\\
- 2\hbar m \omega (1 + \mu_x R_x + \mu_y R_y) \Bigg] \psi(x,y) = \left( \frac{E^2 - m^2 c^4}{c^2} \right) \psi(x,y).\label{hcero}
\end{multline}

Before expressing the differential equation in polar coordinates, we explicitly calculate the commutation relations of the fully deformed non-commutative Dunkl observables to clarify the algebraic structure of this novel phase space.

By substituting the Dunkl momentum $p_{i}^{D}$ into the Seiberg-Witten inverse map (Eqs. (\ref{wm1}) and (\ref{wm2})), and using the fundamental Dunkl commutation relations $[D_x, x] = 1 + 2\mu_x R_x$ and $[D_y, y] = 1 + 2\mu_y R_y$, the coordinate-coordinate and momentum-momentum commutation relations explicitly evaluate to
\begin{align}
[\hat{x}^{D}, \hat{y}^{D}] &= \frac{\theta}{2\hbar}[x, p_{x}^{D}] - \frac{\theta}{2\hbar}[p_{y}^{D}, y] = i\theta(1 + \mu_x R_x + \mu_y R_y), \\
[\hat{p}_{x}^{D}, \hat{p}_{y}^{D}] &= -\frac{\eta}{2\hbar}[p_{x}^{D}, x] + \frac{\eta}{2\hbar}[y, p_{y}^{D}] = i\eta(1 + \mu_x R_x + \mu_y R_y).
\end{align}
Similarly, the coordinate-momentum commutation relations become
\begin{align}
[\hat{x}^{D}, \hat{p}_{x}^{D}] &= [x, p_{x}^{D}] - \frac{\theta\eta}{4\hbar^{2}}[p_{y}^{D}, y] = i\hbar\left(1 + 2\mu_x R_x + \frac{\theta\eta}{4\hbar^{2}}(1 + 2\mu_y R_y)\right), \\
[\hat{y}^{D}, \hat{p}_{y}^{D}] &= [y, p_{y}^{D}] - \frac{\theta\eta}{4\hbar^{2}}[p_{x}^{D}, x] = i\hbar\left(1 + 2\mu_y R_y + \frac{\theta\eta}{4\hbar^{2}}(1 + 2\mu_x R_x)\right),
\end{align}
while the cross-axis commutators vanish $[\hat{x}^{D}, \hat{p}_{y}^{D}] = [\hat{y}^{D}, \hat{p}_{x}^{D}] = 0$.

These explicit relations demonstrate that the non-commutative parameters $\theta$ and $\eta$ are dynamically modified by the discrete parity reflection operators $R_x$ and $R_y$. This proves that our formalism connects phase space deformations with the discrete symmetries of the system.

In polar coordinates, the Dunkl Laplacian takes the form \cite{Genest2013,Genest2015}
\begin{equation}
\nabla_D^2= \frac{\partial^2}{\partial \rho^2}+\frac{1+2\mu_x+2\mu_y}{\rho}\frac{\partial}{\partial \rho}-\frac{2}{\rho^2}B_\phi, \label{laplapol}
\end{equation}
where (as shown in the Appendix) the angular term $B_\phi$ and the Dunkl orbital angular momentum $\mathcal{J}$ are related by the equation
\begin{equation}
\mathcal{J}^2=2B_\phi+2\mu_x\mu_y(1-R_xR_y).
\end{equation}
Using this relation, we can write the Laplacian operator from Eq. (\ref{laplapol}) as
\begin{equation}
\nabla_D^2= \frac{\partial^2}{\partial \rho^2}+\frac{1+2\mu_x+2\mu_y}{\rho}\frac{\partial}{\partial \rho}-\frac{\mathcal{J}^2-2\mu_x\mu_y(1-R_xR_y)}{\rho^2}. \label{laplamomen}
\end{equation}

By substituting this result, Eq. (\ref{hcero}) is expressed in polar coordinates as
\begin{equation}
\begin{split}
\Bigg( -\mathcal{A}\Bigg[\frac{\partial^2}{\partial \rho^2} &+ \frac{1+2\mu_x+2\mu_y}{\rho}\frac{\partial}{\partial \rho}-\frac{\mathcal{J}^2-2\mu_x\mu_y(1-R_xR_y)}{\rho^2}\Bigg] + \mathcal{B} \rho^2 \Bigg) \psi(\rho,\phi) \\
&= \left( \frac{E^2 - m^2 c^4}{c^2}-\mathcal{C} \mathcal{J} + 2\hbar m\omega(1 + \mu_x R_x + \mu_y R_y)\right) \psi(\rho,\phi),\label{completaxx}
\end{split}
\end{equation}
where we have introduced the following constant definitions, which are associated exclusively with the non-commutative phase space and remain completely independent of the Dunkl reflection parameters\begin{eqnarray}
&&\mathcal{A}\equiv   \hbar^2\left( 1 + m^2 \omega^2 \frac{\theta^2}{4\hbar^2} \right), \label{c1}\\
&&\mathcal{B}\equiv   m^2 \omega^2 + \frac{\eta^2}{4\hbar^2}, \label{cb}\\
&&\mathcal{C}\equiv   \eta + m^2 \omega^2 \theta. \label{c2}
\end{eqnarray}

\subsubsection{Commuting Operators and Quantum Numbers}

Before proceeding to the exact analytical solution of the differential equation, it is strictly necessary to identify the Complete Set of Commuting Observables (CSCO) that share simultaneous eigenfunctions with the full Hamiltonian $H_O$. This determines the quantum numbers that consistently label our physical states.

The generalized Hamiltonian of Eq. (\ref{completaxx}), in terms of the definitions (\ref{c1})-(\ref{c2}), is given by
\begin{equation}
H_{O} = -\mathcal{A}\nabla_{D}^{2} + \mathcal{B}\rho^{2} + \mathcal{C}\mathcal{J} - 2m\omega\hbar(1+\mu_x R_x + \mu_y R_y).
\end{equation}
We must analyze the commutation relations of $H_O$ with the spatial reflection operators $R_x$ and $R_y$, and the Dunkl angular momentum $\mathcal{J}=i(xD_y - yD_x)$. Both the Dunkl Laplacian $\nabla_D^2$ and the radial variable $\rho^2$ are invariant under spatial reflections, meaning they commute with $R_x$, $R_y$, and $\mathcal{J}$. However, the Dunkl angular momentum $\mathcal{J}$ satisfies the anti-commutation relations
\begin{align}
R_x \mathcal{J} &= -\mathcal{J} R_x, \\
R_y \mathcal{J} &= -\mathcal{J} R_y.
\end{align}
Because $\mathcal{J}$ anti-commutes with $R_x$ and $R_y$, we have $[R_x, \mathcal{J}] \neq 0$ and $[R_y, \mathcal{J}] \neq 0$. Since our full Hamiltonian $H_O$ contains a linear term in $\mathcal{J}$, the Hamiltonian does not commute with the individual reflection operators: $[R_x, H_O] \neq 0$ and $[R_y, H_O] \neq 0$. Consequently, $s_x$ and $s_y$ do not constitute valid quantum numbers for characterizing the physical states of the system.

Nevertheless, the product of the reflection operators, $R_x R_y$, restores the commutation due to the double sign cancellation
\begin{equation}
(R_x R_y)\mathcal{J} = R_x(-\mathcal{J}R_y) = \mathcal{J}(R_x R_y) \implies [R_x R_y, \mathcal{J}] = 0.
\end{equation}
Therefore, the parity operator $R_x R_y$ commutes with both $\mathcal{J}$ and $H_O$. The complete set of mutually commuting operators is given by $\{ H_O, \mathcal{J}, R_x R_y \}$. The eigenstates of the system $\psi(\rho, \phi)$ are labeled by the following set of quantum numbers:
\begin{enumerate}
    \item Radial quantum number ($k$): As we will see in Section 2.4, this number forces the power series to truncate into the associated Laguerre polynomials, and is restricted to non-negative integers, $k \in \mathbb{N} \cup \{0\}$.
    \item Reflection eigenvalues ($s_x, s_y$): As shown in the Appendix, these are the eigenvalues of the spatial reflection operators, which can only take the values $s_x \in \{+1, -1\}$ and $s_y \in \{+1, -1\}$. Their product defines the joint parity quantum number $\epsilon = s_x s_y \in \{+1, -1\}$.
    \item Angular quantum number ($\ell$): As detailed in the Appendix, this number defines the Dunkl angular momentum eigenvalue $\lambda_\epsilon$. Its allowed range is determined by the parity sector $\epsilon$:
    \begin{itemize}
        \item If $\epsilon = 1$ (aligned parity, $s_x = s_y$), then $\ell \in \mathbb{N} \cup \{0\}$.
        \item If $\epsilon = -1$ (anti-aligned parity, $s_x = -s_y$), then $\ell \in \{1/2, 3/2, 5/2, \dots\}$.
    \end{itemize}
\end{enumerate}

\subsection{Exact solutions for the Dunkl-Klein-Gordon oscillator}

We use the mathematical results detailed in the Appendix to solve the radial differential equation. The angular eigenfunctions satisfy the eigenvalue equation
\begin{equation}
\mathcal{J}F_{\epsilon}=\lambda_{\epsilon}F_{\epsilon},
\end{equation}
where $s_{x}$ and $s_{y}$ are the eigenvalues of the reflection operators $R_{x}$ and $R_{y}$, while $\epsilon\equiv s_{x}s_{y}=\pm 1$. The eigenfunctions and eigenvalues of the operator $\mathcal{J}$ are classified into two distinct cases.

\begin{itemize}
\item First, the condition $R_{x}=R_{y}$ implies $\epsilon=1$. In this case, the eigenvalue equation has the solutions
\begin{align}
F_{+} &= \Phi_{\ell}^{++}(\phi)\pm i\Phi_{\ell}^{--}(\phi), \\
\lambda_{+} &= \pm 2\sqrt{\ell(\ell+\mu_{x}+\mu_{y})},
\end{align}
where $\ell\in\mathbb{N}$, and the specific forms of the functions $\Phi_{\ell}^{++}$ and $\Phi_{\ell}^{--}$ are provided in the Appendix. This scenario occurs when $(s_{x},s_{y})=(1,1)$ or $(s_{x},s_{y})=(-1,-1)$. By defining the wave functions as $\psi_{\pm\pm}^{+}=\psi_{\pm\pm}^{+}(\rho)F_{+}(\phi)$, the differential equation is rewritten as
\begin{equation}
\begin{split}
\Bigg( -\mathcal{A}\Bigg[\frac{\partial^{2}}{\partial\rho^{2}} &+ \frac{1+2\mu_{x}+2\mu_{y}}{\rho}\frac{\partial}{\partial\rho}-\frac{\lambda_{+}^{2}}{\rho^{2}}\Bigg]+\mathcal{B}\rho^{2}\Bigg)\psi_{\pm\pm}^{+}(\rho) \\
&= \left(\frac{(E_{\pm\pm}^{+})^{2}-m^{2}c^{4}}{c^{2}}-\mathcal{C}\lambda_{+}+ 2m\omega\hbar(1 \pm \mu_{x} \pm \mu_{y})\right)\psi_{\pm\pm}^{+}(\rho).
\end{split}
\end{equation}
By solving the radial differential equation and imposing the physical requirement that the wave functions must remain finite at the origin, we determine the regular solutions in terms of the associated Laguerre polynomials
\begin{equation}
\psi_{\pm\pm}^{+}(\rho)=\rho^{\sqrt{\lambda_{+}^{2}+(\mu_{x}+\mu_{y})^{2}}-\mu_{x}-\mu_{y}}e^{-\frac{1}{2}\sqrt{\frac{\mathcal{B}}{\mathcal{A}}}\rho^{2}}L_{k}^{\sqrt{\lambda_{+}^{2}+(\mu_{x}+\mu_{y})^{2}}}\left(\sqrt{\frac{\mathcal{B}}{\mathcal{A}}}\rho^{2}\right).
\end{equation}
To ensure that these wave functions do not diverge at $\rho\rightarrow\infty$, the polynomial series must truncate. This requires the parameter $k$ to be a non-negative integer, establishing the fundamental quantization condition
{\small\begin{equation}
\frac{1}{4\sqrt{\mathcal{B}}\sqrt{\mathcal{A}}}\left(\frac{(E_{\pm\pm}^{+})^{2}-m^{2}c^{4}}{c^{2}}-\mathcal{C}\lambda_{+}+2m\omega\hbar(1 \pm \mu_{x} \pm \mu_{y})\right)-\frac{1}{2}\sqrt{\lambda_{+}^{2}+(\mu_{x}+\mu_{y})^{2}}-\frac{1}{2}=k.
\end{equation}}
Solving this relation algebraically yields the discrete energy levels of the system
{\small\begin{equation}
E_{\pm\pm}^{+}=\pm mc^{2}\sqrt{1+\frac{2\sqrt{\mathcal{B}}\sqrt{\mathcal{A}}\left(\sqrt{\lambda_{+}^{2}+(\mu_{x}+\mu_{y})^{2}}+1+2k\right)+\mathcal{C}\lambda_{+}-2m\omega\hbar(1 \pm \mu_{x} \pm \mu_{y})}{m^{2}c^{2}}}. \label{espec1}
\end{equation}}
Here, the upper signs inside the parity term correspond to the state $(s_{x},s_{y})=(1,1)$, while the lower signs correspond to $(-1,-1)$.

\item Second, the condition $R_{x}=-R_{y}$ implies $\epsilon=-1$. This is satisfied when $(s_{x},s_{y})=(1,-1)$ or $(s_{x},s_{y})=(-1,1)$. The eigenfunctions and eigenvalues of the angular momentum take the forms
\begin{align}
F_{-} &= \Phi_{\ell}^{-+}(\phi)\mp i\Phi_{\ell}^{+-}(\phi), \\
\lambda_{-} &= \pm 2\sqrt{(\ell+\mu_{x})(\ell+\mu_{y})},
\end{align}
where $\ell\in\{\frac{1}{2},\frac{3}{2},...\}$. By defining the wave functions as $\psi_{\pm\mp}^{-}=\psi_{\pm\mp}^{-}(\rho)F_{-}(\phi)$, the differential equation becomes
\begin{equation}
\begin{split}
\Bigg(-\mathcal{A}\Bigg[\frac{\partial^{2}}{\partial\rho^{2}} &+ \frac{1+2\mu_{x}+2\mu_{y}}{\rho}\frac{\partial}{\partial\rho}-\frac{\lambda_{-}^{2}-4\mu_{x}\mu_{y}}{\rho^{2}}\Bigg]+\mathcal{B}\rho^{2}\Bigg)\psi_{\pm\mp}^{-}(\rho) \\
&= \left(\frac{(E_{\pm\mp}^{-})^{2}-m^{2}c^{4}}{c^{2}}-\mathcal{C}\lambda_{-}+2m\omega\hbar(1 \pm \mu_{x} \mp \mu_{y})\right)\psi_{\pm\mp}^{-}(\rho).
\end{split}
\end{equation}
Demanding the radial functions to be well-behaved at the origin yields the regular solutions
\begin{equation}
\psi_{\pm\mp}^{-}(\rho)=\rho^{\sqrt{\lambda_{-}^{2}+(\mu_{x}-\mu_{y})^{2}}-\mu_{x}-\mu_{y}}e^{-\frac{1}{2}\sqrt{\frac{\mathcal{B}}{\mathcal{A}}}\rho^{2}}L_{k}^{\sqrt{\lambda_{-}^{2}+(\mu_{x}-\mu_{y})^{2}}}\left(\sqrt{\frac{\mathcal{B}}{\mathcal{A}}}\rho^{2}\right).
\end{equation}
The boundary condition at $\rho\rightarrow\infty$ forces the series to terminate ($k \in \mathbb{N} \cup \{0\}$), leading to the quantization relation
{\small\begin{equation}
\frac{1}{4\sqrt{\mathcal{B}}\sqrt{\mathcal{A}}}\left(\frac{(E_{\pm\mp}^{-})^{2}-m^{2}c^{4}}{c^{2}}-\mathcal{C}\lambda_{-}+2m\omega\hbar(1 \pm \mu_{x} \mp \mu_{y})\right)-\frac{1}{2}\sqrt{\lambda_{-}^{2}+(\mu_{x}-\mu_{y})^{2}}-\frac{1}{2}=k.
\end{equation}}
Solving for the energy, the exact spectrum for this parity configuration is expressed as
{\small\begin{equation}
E_{\pm\mp}^{-}=\pm mc^{2}\sqrt{1+\frac{2\sqrt{\mathcal{B}}\sqrt{\mathcal{A}}\left(\sqrt{\lambda_{-}^{2}+(\mu_{x}-\mu_{y})^{2}}+1+2k\right)+\mathcal{C}\lambda_{-}-2m\omega\hbar(1 \pm \mu_{x} \mp \mu_{y})}{m^{2}c^{2}}}. \label{espec2}
\end{equation}}
For this final expression, the upper signs correspond to the $(s_{x},s_{y})=(1,-1)$ state, whereas the lower signs describe the $(s_{x},s_{y})=(-1,1)$ case.
\end{itemize}

\subsection{Commutative Limits}

In order to verify the physical consistency of the exact generalized spectra obtained via the sequential deformation, we analytically obtain the spectral behavior when the non-commutative phase space deformations vanish ($\theta \rightarrow 0$ and $\eta \rightarrow 0$).

Taking this commutative limit, the scale parameters defined previously reduce to
\begin{equation}
\mathcal{A} \rightarrow \hbar^2, \quad \mathcal{B} \rightarrow m^2\omega^2, \quad \mathcal{C} \rightarrow 0.
\end{equation}
Consequently, the factor $2\sqrt{\mathcal{B}}\sqrt{\mathcal{A}}$ reduces to $2\hbar m\omega$. Substituting these limits into the exact energy spectrum for the aligned parity states ($E_{\pm\pm}^{+}$) and simplifying the inner square root corresponding to $\sqrt{4\ell(\ell+\mu_x+\mu_y) + (\mu_x+\mu_y)^2} = 2\ell + \mu_x + \mu_y$, the energy spectrum simplifies to
\begin{equation}
E_{\pm\pm}^{+} = \pm mc^{2}\sqrt{1+\frac{2\hbar m\omega\left(2\ell + \mu_x + \mu_y + 1 + 2k\right) - 2m\omega\hbar(1 \pm \mu_x \pm \mu_y)}{m^{2}c^{2}}}.
\end{equation}
By factoring the terms inside the parenthesis, we obtain the following expression
\begin{equation}
E_{\pm\pm}^{+} = \pm mc^{2}\sqrt{1+\frac{4\hbar\omega}{mc^{2}}\left(k + \ell + \frac{\mu_x(1\mp 1)}{2} + \frac{\mu_y(1\mp 1)}{2} \right)}. \label{limit1}
\end{equation}

Applying the identical commutative limit to the anti-aligned parity states ($E_{\pm\mp}^{-}$), the inner root simplifies to $\sqrt{\lambda_{-}^2+(\mu_x-\mu_y)^2} = 2\ell + \mu_x + \mu_y$. This result leads us to obtain
\begin{equation}
E_{\pm\mp}^{-} = \pm mc^{2}\sqrt{1+\frac{4\hbar\omega}{mc^{2}}\left(k + \ell + \frac{\mu_x(1\mp 1)}{2} + \frac{\mu_y(1\pm 1)}{2} \right)}. \label{limit2}
\end{equation}

Eqs. (\ref{limit1}) and (\ref{limit2}) exactly reproduce the energy spectra computed for the standard commutative Dunkl-Klein-Gordon oscillator reported in Ref. \cite{Nos2021}, with $k$ playing the role of the principal quantum number. This reduction proves that the sequential ordering of the deformations (Dunkl derivative followed by the Seiberg-Witten map) preserves the discrete symmetries of the Dunkl reflections while correcting the phase space couplings.

\section{Dunkl-Klein-Gordon oscillator in a uniform magnetic field}

We start from the standard Klein-Gordon oscillator given by Eq. (\ref{kgo}) and introduce a uniform magnetic field along the $z$-direction through minimal coupling. Applying this procedure yields the equation
\begin{equation}
\left[ \left(\mathbf{p} - \frac{q}{c}\mathbf{A}\right) + i m \omega \boldsymbol{\rho}\right] \cdot \left[ \left(\mathbf{p} - \frac{q}{c}\mathbf{A}\right) - i m \omega \boldsymbol{\rho} \right] \psi(\boldsymbol{\rho}) = \left( \frac{E^2 - m^2 c^4}{c^2} \right) \psi(\boldsymbol{\rho}).
\end{equation}
We now introduce the vector potential in the symmetric gauge $\mathbf{A} = \frac{B}{2} (-y, x)$, which satisfies $\nabla\times \mathbf{A}= B\hat{k}$. Assuming the oscillator carries the elementary charge of an electron, we replace the general coupling factor $-q$ with the absolute value $e$. Incorporating these terms, the equation takes the explicit form
\begin{equation}
\left[ \left(\mathbf{p} + \frac{e B}{2c} (-y, x)\right) + i m \omega \boldsymbol{\rho} \right] \cdot \left[ \left(\mathbf{p} + \frac{e B}{2c} (-y,x)\right) - i m \omega \boldsymbol{\rho} \right] \psi(\boldsymbol{\rho}) = \left( \frac{E^2 - m^2 c^4}{c^2} \right) \psi(\boldsymbol{\rho}).
\end{equation}
To preserve the logical sequence established in Section 2, we define the operator $\mathbf{\Pi} = \mathbf{p} + \frac{eB}{2c}(-y,x)$ and introduce the Dunkl derivative into the commutative space, $\mathbf{p} \rightarrow \mathbf{p}^D$. Expanding the left-hand side and using the canonical commutation relation of the Dunkl derivative, we obtain $-im\omega(p_x^D x - x p_x^D + p_y^D y - y p_y^D) = -2m\omega\hbar(1+\mu_x R_x + \mu_y R_y)$. Thus, the differential equation in the commutative Dunkl space is
\begin{multline}
\left[ (p^D)^2 + \frac{eB}{c}L_z^D + \left( \frac{e^2 B^2}{4c^2} + m^2 \omega^2 \right) \rho^2 - 2 m \omega \hbar(1+\mu_x R_x + \mu_y R_y) \right] \psi(\boldsymbol{\rho})\\
 = \left( \frac{E^2 - m^2 c^4}{c^2} \right) \psi(\boldsymbol{\rho}).
\end{multline}
This result represents the Klein-Gordon oscillator coupled to a uniform magnetic field within the Dunkl formalism in a commutative space. We use this equation as the starting point to study the non-commutative regime.

\subsection{Dunkl-Klein-Gordon oscillator in a uniform magnetic field in the non-commutative phase space}

We extend our problem to a non-commutative phase space by introducing the non-commutative coordinates and momenta into the generalized Klein-Gordon oscillator equation. Thus, we obtain
\begin{multline}
\bigg[ (\hat{p}^D)^2 + \frac{eB}{c}\hat{L}_z^D + \left( \frac{e^2 B^2}{4c^2} + m^2 \omega^2 \right) (\hat{\rho}^{D})^2 \\
 - 2 m \omega \hbar (1 + \mu_x R_x + \mu_y R_y) \bigg] \star \psi(x,y) = \left( \frac{E^2 - m^2 c^4}{c^2} \right) \psi(x,y).
\end{multline}

We map this non-commutative equation back to the commutative space using the Seiberg-Witten transformations defined in Eqs. (\ref{wm1}) and (\ref{wm2}). Applying this map and simplifying the algebraic terms, we find
{\small
\begin{equation}
\begin{split}
\Bigg[ -\hbar^2\left( 1 - \frac{eB\theta}{2\hbar c} + \left( \frac{e^2 B^2}{4c^2} + m^2 \omega^2 \right) \frac{\theta^2}{4\hbar^2} \right) \nabla_D^2 &+ \left( \frac{\eta^2}{4\hbar^2} - \frac{eB\eta}{2\hbar c} + \left( \frac{e^2 B^2}{4c^2} + m^2 \omega^2 \right) \right) \rho^2 \\
+ \left( \eta - \frac{eB\hbar}{c}\left( 1 + \frac{\theta\eta}{4\hbar^2} \right) + \left( \frac{e^2 B^2}{4c^2} + m^2 \omega^2 \right) \theta \right) \mathcal{J} &- 2m\omega\hbar(1+\mu_x R_x + \mu_y R_y) \Bigg]\psi(\rho,\phi)=\left(\frac{E^2-m^2c^4}{c^2}\right)\psi(\rho,\phi).
\end{split}
\end{equation}
}
This allows us to write the differential equation in the compact polar form
\begin{equation}
\left( -\tilde{\mathcal{A}} \nabla_D^2 + \tilde{\mathcal{B}} \rho^2 + \tilde{\mathcal{C}} \mathcal{J} - 2m\omega\hbar(1+\mu_x R_x + \mu_y R_y) \right) \psi(\rho,\phi) = \left( \frac{E^2 - m^2 c^4}{c^2} \right) \psi(\rho,\phi), \label{dkgf}
\end{equation}
where the Dunkl Laplacian $\nabla_D^2$ is given explicitly by Eq. (\ref{laplamomen}), and the phase space constants are defined by
\begin{eqnarray}
&&\tilde{\mathcal{A}}\equiv \hbar^2\left( 1 - \frac{eB\theta}{2\hbar c} + \left( \frac{e^2 B^2}{4c^2} + m^2 \omega^2 \right) \frac{\theta^2}{4\hbar^2} \right), \label{c3}\\
&&\tilde{\mathcal{B}}\equiv  \frac{\eta^2}{4\hbar^2} - \frac{eB\eta}{2\hbar c} + \left( \frac{e^2 B^2}{4c^2} + m^2 \omega^2 \right), \label{c3b}\\
&&\tilde{\mathcal{C}}\equiv   \eta - \frac{eB\hbar}{c}\left( 1 + \frac{\theta\eta}{4\hbar^2} \right) + \left( \frac{e^2 B^2}{4c^2} + m^2 \omega^2 \right) \theta. \label{c4}
\end{eqnarray}
To solve Eq. (\ref{dkgf}), we take advantage of its exact mathematical similarity to the field-free case. This structural equivalence enables us to directly map the analytical solutions derived in the previous section to the present case.

Consequently, the regular wave functions at the origin and the generalized Landau energy levels for the parity states $(s_x,s_y)=(1,1)$ and $(s_x,s_y)=(-1,-1)$ take the form
{\small
\begin{eqnarray}
&&\psi_{\pm\pm}^+(\rho)=\rho^{\sqrt{\lambda_+^2+(\mu_x+\mu_y)^2}-\mu_x-\mu_y} e^{-\frac{1}{2}\sqrt{\frac{\tilde{\mathcal{B}}}{\tilde{\mathcal{A}}}}\rho^2}L_{k}^{\sqrt{\lambda_+^2+(\mu_x+\mu_y)^2}}\left(\sqrt{\tilde{\mathcal{B}}/{\tilde{\mathcal{A}}}}\rho^2\right)\label{f1CM},\\
&&E_{\pm\pm}^+ = \pm mc^2 \sqrt{1 + \frac{2\sqrt{\tilde{\mathcal{B}}} \sqrt{\tilde{\mathcal{A}}} \left(\sqrt {\lambda_+^2+(\mu_x+\mu_y)^2} + 1+ 2k \right) + \tilde{\mathcal{C}}\lambda_+ - 2m\omega\hbar(1 \pm\mu_x \pm \mu_y)}{m^2c^2}}.\label{E1CM}
\end{eqnarray}}
Here, the upper signs in the term $1 \pm\mu_x \pm \mu_y$ correspond to the parity state $(s_x, s_y)=(1,1)$, while the lower signs correspond to the state $(s_x, s_y)=(-1,-1)$.

Following exactly the same procedure for the anti-aligned parity cases $(s_x,s_y)=(1,-1)$ and $(s_x,s_y)=(-1,1)$, the regular wave functions and their associated energy spectra are expressed as
{\small
\begin{eqnarray}
&&\psi_{\pm\mp}^-(\rho)=\rho^{\sqrt{\lambda_-^2+(\mu_x-\mu_y)^2}-\mu_x-\mu_y} e^{-\frac{1}{2}\sqrt{\frac{\tilde{\mathcal{B}}}{\tilde{\mathcal{A}}}}\rho^2}L_{k}^{\sqrt{\lambda_-^2+(\mu_x-\mu_y)^2}}\left(\sqrt{{\tilde{\mathcal{B}}}/{\tilde{\mathcal{A}}}}\rho^2\right),\label{f2CM}\\
&&E_{\pm\mp}^- = \pm mc^2 \sqrt{1 + \frac{2\sqrt{\tilde{\mathcal{B}}} \sqrt{\tilde{\mathcal{A}}} \left(\sqrt {\lambda_-^2+(\mu_x-\mu_y)^2} + 1+ 2k \right) + \tilde{\mathcal{C}}\lambda_- - 2m\omega\hbar(1 \pm\mu_x \mp \mu_y)}{m^2c^2}}.\label{E2CM}
\end{eqnarray}}
In this last result, the upper signs correspond to the state $(s_x, s_y)=(1,-1)$, and the lower signs correspond to the state $(s_x, s_y)=(-1,1)$.

\subsection{Special Limiting Cases}

In order to verify the physical consistency of our unified framework, we demonstrate the analytical reduction of the derived generalized Landau levels under specific physical and geometric limits.
\begin{enumerate}
\item Vanishing Magnetic Field ($B \rightarrow 0$):
By turning off the external magnetic field, the cyclotron frequency $\Omega$ tends to the harmonic oscillator frequency $\omega$, and the scale parameters defined in Eqs. (\ref{c3})--(\ref{c4}) reduce to the unmagnetized phase space constants: $\tilde{\mathcal{A}} \rightarrow \mathcal{A}$, $\tilde{\mathcal{B}} \rightarrow \mathcal{B}$, and $\tilde{\mathcal{C}} \rightarrow \mathcal{C}$. Consequently, the generalized Landau levels in Eqs. (\ref{E1CM}) and (\ref{E2CM}) reduce to the non-commutative Dunkl-Klein-Gordon oscillator energy levels of Eqs. (\ref{espec1}) and (\ref{espec2}), respectively, derived in Section 2.

\item Commutative Dunkl-Landau Levels ($\theta \rightarrow 0, \eta \rightarrow 0$):
When the non-commutative phase space deformations vanish, the effective scale parameters reduce to
\begin{equation}
\tilde{\mathcal{A}} \rightarrow \hbar^2, \quad \tilde{\mathcal{B}} \rightarrow m^2\omega^2 + \frac{e^2B^2}{4c^2} \equiv m^2\Omega^2, \quad \tilde{\mathcal{C}} \rightarrow -\frac{eB\hbar}{c}.
\end{equation}
Substituting these values into the generalized Landau levels of Eq. (\ref{E1CM}) for the aligned parity states ($E_{\pm\pm}^{+}$) and simplifying the inner square root $\sqrt{4\ell(\ell+\mu_x+\mu_y) + (\mu_x+\mu_y)^2} = 2\ell + \mu_x + \mu_y$, the energy spectrum takes the form
{\small
\begin{equation}
E_{\pm\pm}^{+} = \pm mc^{2}\sqrt{1+\frac{2\hbar m\Omega\left(2\ell + \mu_x + \mu_y + 1 + 2k\right) - \frac{eB\hbar}{c}\lambda_{+} - 2m\omega\hbar(1 \pm \mu_x \pm \mu_y)}{m^{2}c^{2}}}.
\end{equation}}
By factoring the parameters $\mu_x$ and $\mu_y$, we obtain the unified expression
{\small\begin{equation}
E_{\pm\pm}^{+} = \pm mc^{2}\sqrt{1+\frac{4\hbar\Omega}{mc^{2}}\left(k + \ell + \frac{1}{2}\left(1-\frac{\omega}{\Omega}\right) + \frac{\mu_x}{2}\left(1 \mp \frac{\omega}{\Omega}\right) + \frac{\mu_y}{2}\left(1 \mp \frac{\omega}{\Omega}\right) - \frac{eB}{4m\Omega c}\lambda_{+}\right)}. \label{limit3}
\end{equation}}

Applying a similar procedure to the anti-aligned parity states spectrum ($E_{\pm\mp}^{-}$), where $\sqrt{\lambda_{-}^2+(\mu_x-\mu_y)^2} = 2\ell + \mu_x + \mu_y$, we obtain
{\small\begin{equation}
E_{\pm\mp}^{-} = \pm mc^{2}\sqrt{1+\frac{4\hbar\Omega}{mc^{2}}\left(k + \ell + \frac{1}{2}\left(1-\frac{\omega}{\Omega}\right) + \frac{\mu_x}{2}\left(1 \mp \frac{\omega}{\Omega}\right) +\frac{\mu_y}{2}\left(1 \pm \frac{\omega}{\Omega}\right) - \frac{eB}{4m\Omega c}\lambda_{-}\right)}. \label{limit4}
\end{equation}}
Eqs. (\ref{limit3}) and (\ref{limit4}) exactly reproduce the generalized commutative Landau levels reported in Ref. \cite{Nos2021b}.

\item Standard Commutative Dunkl-Klein-Gordon Oscillator ($\theta, \eta, B \rightarrow 0$):
By turning off the magnetic field ($B \rightarrow 0$) within the commutative limit, the magnetic coupling term vanishes ($\frac{eB}{4m\Omega c}\lambda_{\pm} \rightarrow 0$) and $\Omega \rightarrow \omega$ (hence $\frac{\omega}{\Omega} \rightarrow 1$). The spectra in Eqs. (\ref{limit3}) and (\ref{limit4}) reduce to
\begin{align}
E_{\pm\pm}^{+} &= \pm mc^{2}\sqrt{1+\frac{4\hbar\omega}{mc^{2}}\left(k + \ell + \frac{\mu_x(1\mp 1)}{2} +\frac{\mu_y(1\mp 1)}{2}\right)}, \\
E_{\pm\mp}^{-} &= \pm mc^{2}\sqrt{1+\frac{4\hbar\omega}{mc^{2}}\left(k + \ell + \frac{\mu_x(1\mp 1)}{2} +\frac{\mu_y(1\pm 1)}{2}\right)},
\end{align}
which recover the spectra of the generalized Klein-Gordon equation with Dunkl derivative reported in Ref. \cite{Nos2021}. These explicit limits demonstrate that generalizing the system first with the Dunkl derivative and then into a non-commutative phase space preserves both the parity effects of the Dunkl reflections and the standard behavior of the Klein-Gordon oscillator.
\end{enumerate}

\subsection{Non-relativistic limit}

In order to explore the low-energy regime of our generalized system, we explicitly compute its non-relativistic limit ($c \to \infty$). To extract this limit for the positive-energy branch, we define the total relativistic energy as $E = mc^2 + \mathcal{E}_{NR}$, where $mc^2$ is the rest mass energy and $\mathcal{E}_{NR}$ is the non-relativistic energy.

In the limit $mc^2 \gg \mathcal{E}_{NR}$, the kinetic term on the right-hand side of our differential equation Eq. (\ref{dkgf}) expands as
\begin{equation}
\frac{E^2 - m^2 c^4}{c^2} = \frac{(mc^2 + \mathcal{E}_{NR})^2 - m^2 c^4}{c^2} = 2m\mathcal{E}_{NR} + \frac{\mathcal{E}_{NR}^2}{c^2} \approx 2m\mathcal{E}_{NR}.
\end{equation}
By substituting this approximation back into Eq. (\ref{dkgf}) and dividing by $2m$, we obtain the differential equation
\begin{equation}
\left( -\frac{\tilde{\mathcal{A}}}{2m} \nabla_D^2 + \frac{\tilde{\mathcal{B}}}{2m} \rho^2 + \frac{\tilde{\mathcal{C}}}{2m} \mathcal{J} - \omega\hbar(1+\mu_x R_x + \mu_y R_y) \right) \psi(\rho,\phi) = \mathcal{E}_{NR} \psi(\rho,\phi). \label{SchrodingerNC}
\end{equation}
This resulting expression represents the non-relativistic Schr\"odinger-Dunkl oscillator equation in a non-commutative phase space coupled to a uniform magnetic field.

Consequently, to find the non-relativistic energy spectrum, we apply the same limit $c \to \infty$ directly to our exact generalized Landau levels. Taking the positive root of Eq. (\ref{E1CM}), we factor $mc^2$ out of the square root and apply a first-order Taylor expansion $\sqrt{1 + u} \approx 1 + \frac{u}{2}$
\begin{equation}
E \approx mc^2 \left( 1 + \frac{2\sqrt{\tilde{\mathcal{B}}} \sqrt{\tilde{\mathcal{A}}} \left(\sqrt {\lambda_+^2+(\mu_x+\mu_y)^2} + 1+ 2k \right) + \tilde{\mathcal{C}}\lambda_+ - 2m\omega\hbar(1 \pm\mu_x \pm \mu_y)}{2m^2c^2} \right).
\end{equation}
Subtracting the rest mass energy, we obtain the non-relativistic energy spectrum for the $(1,1)$ and $(-1,-1)$ parity sectors
\begin{equation}
\mathcal{E}_{NR}^{+} = \frac{\sqrt{\tilde{\mathcal{B}}\tilde{\mathcal{A}}}}{m} \left(\sqrt {\lambda_+^2+(\mu_x+\mu_y)^2} + 1+ 2k \right) + \frac{\tilde{\mathcal{C}}}{2m}\lambda_+ - \omega\hbar(1 \pm\mu_x \pm \mu_y).
\end{equation}
Similarly, for the anti-aligned parity states $(1,-1)$ and $(-1,1)$, the non-relativistic limit evaluates to
\begin{equation}
\mathcal{E}_{NR}^{-} = \frac{\sqrt{\tilde{\mathcal{B}}\tilde{\mathcal{A}}}}{m} \left(\sqrt {\lambda_-^2+(\mu_x-\mu_y)^2} + 1+ 2k \right) + \frac{\tilde{\mathcal{C}}}{2m}\lambda_- - \omega\hbar(1 \pm\mu_x \mp \mu_y).
\end{equation}
These results demonstrate that both the non-commutative phase space geometry parameters ($\theta, \eta$) and the Dunkl parity deformations ($\mu_x, \mu_y$) survive in the non-relativistic limit. The non-relativistic limits for the unmagnetized case discussed in Section 2 are recovered by evaluating these expressions at $B=0$. It is worth emphasizing that the differential equation Eq. (\ref{SchrodingerNC}) and the non-relativistic energy spectra $\mathcal{E}_{NR}^{\pm}$, to our knowledge, have not been previously reported in the literature.

\subsection{Graphical Analysis of the Energy Spectrum}

To fully understand the physical implications of our generalized model, we graphically analyze the positive-energy branch of the exact spectra given by Eqs. (\ref{E1CM}) and (\ref{E2CM}). For all the following representations, we adopt natural units ($m=c=\hbar=\omega=e=1$) and study the fundamental state ($k=0$) to isolate the pure effects of the non-commutative and Dunkl parameters.

\begin{figure}[hbt!]
    \centering
    \includegraphics[width=0.8\textwidth]{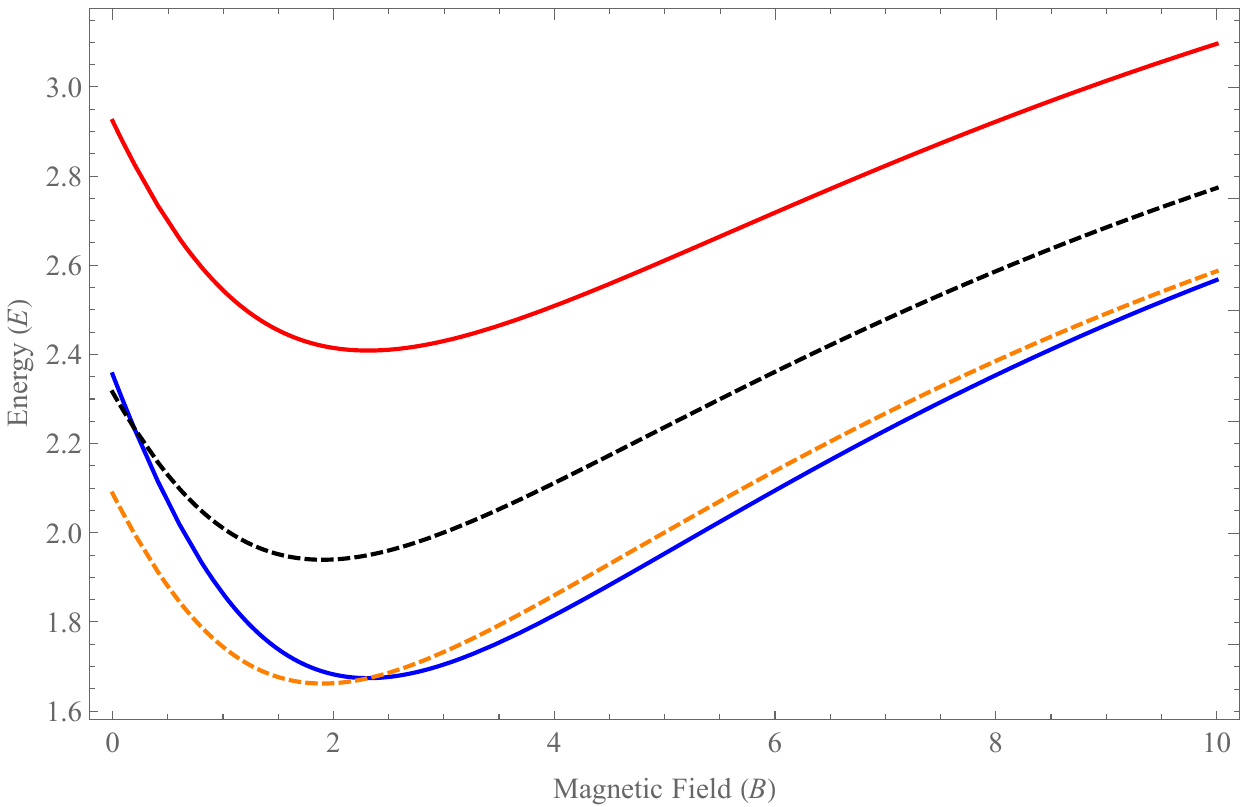}
    \caption{Energy spectrum $E$ as a function of the external magnetic field $B$. The non-commutative geometric parameters are fixed at $\theta=0.1$ and $\eta=0.1$, while the Dunkl parameters are set to constant values of $\mu_x=0.5$ and $\mu_y=0.25$. Solid lines represent integer angular momentum states ($\ell=1$) for the aligned parity $(1,1)$ in blue, and $(-1,-1)$ in red. Dashed lines represent half-integer angular momentum states ($\ell=1/2$) for the anti-aligned parities $(1,-1)$ in orange, and $(-1,1)$ in black.}
    \label{fig:magnetic}
\end{figure}

In Figure \ref{fig:magnetic}, we observe the evolution of the energy levels as the uniform magnetic field $B$ increases. The graph shows that the energy curves for the different parity sectors are clearly separated into distinct branches. For weak magnetic fields, the energy initially decreases until it reaches a minimum value. As the magnetic field grows stronger, the energy increases linearly. This behavior illustrates how the combination of the Dunkl parameters and the non-commutative phase space creates distinct, separated energy paths for each state.

\begin{figure}[hbt!]
    \centering
    \includegraphics[width=0.8\textwidth]{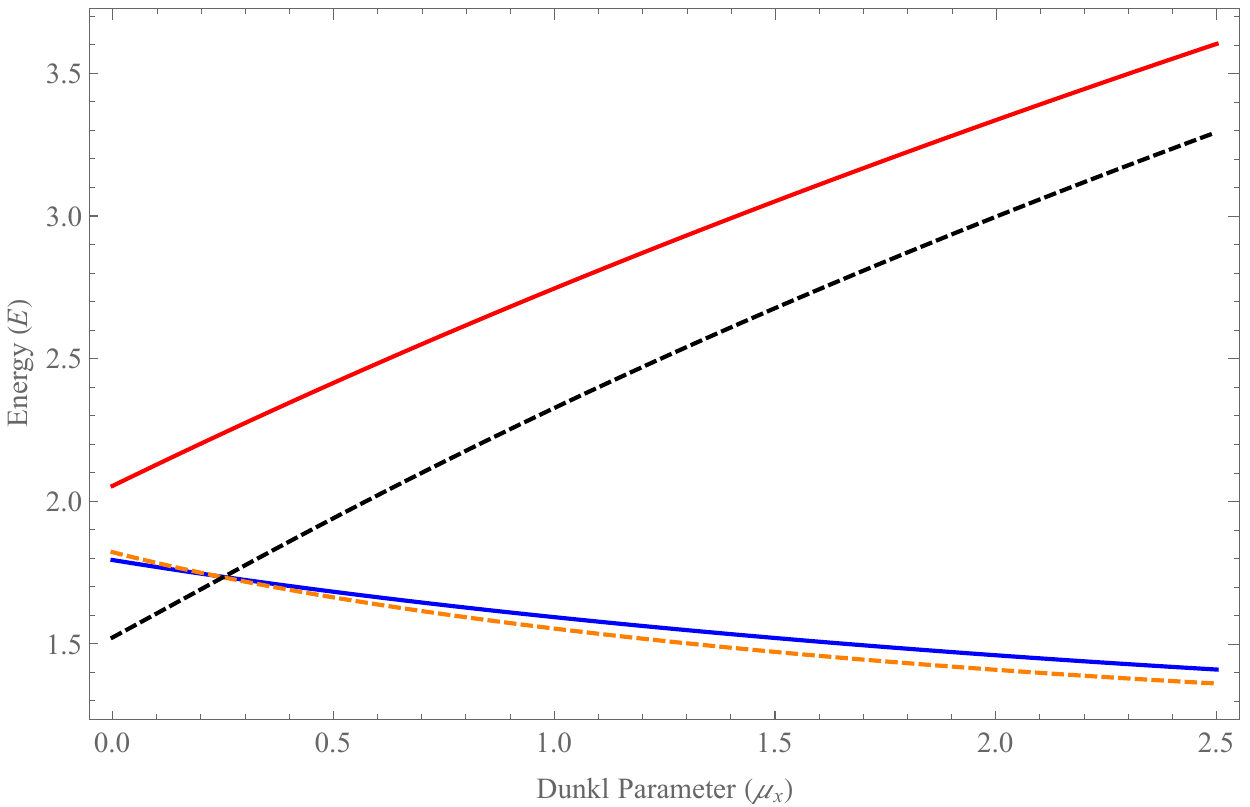}
    \caption{Energy spectrum $E$ as a function of the Dunkl parameter $\mu_x$. The constant values are set to $B=2$, $\theta=0.1$, $\eta=0.1$, and $\mu_y=0.25$. Following the color scheme, solid lines correspond to $\ell=1$ states with parity $(1,1)$ (blue) and $(-1,-1)$ (red). Dashed lines correspond to $\ell=1/2$ states with parity $(1,-1)$ (orange) and $(-1,1)$ (black).}
    \label{fig:mux}
\end{figure}

Figure \ref{fig:mux} isolates the effect of the Dunkl parameter $\mu_x$. As $\mu_x$ increases, the energy of the $(-1,-1)$ parity state (red solid line) rises significantly, whereas the $(1,1)$ parity state (blue solid line) experiences a slight decrease in energy. Similarly, for the anti-aligned sectors, the $(-1,1)$ state (black dashed line) increases with $\mu_x$, while the $(1,-1)$ state (orange dashed line) decreases slightly. This explicit asymmetry proves that the spatial reflection symmetries modify the dynamic behavior of the scalar bosons depending exclusively on their specific parity configuration.

\begin{figure}[hbt!]
    \centering
    \includegraphics[width=0.8\textwidth]{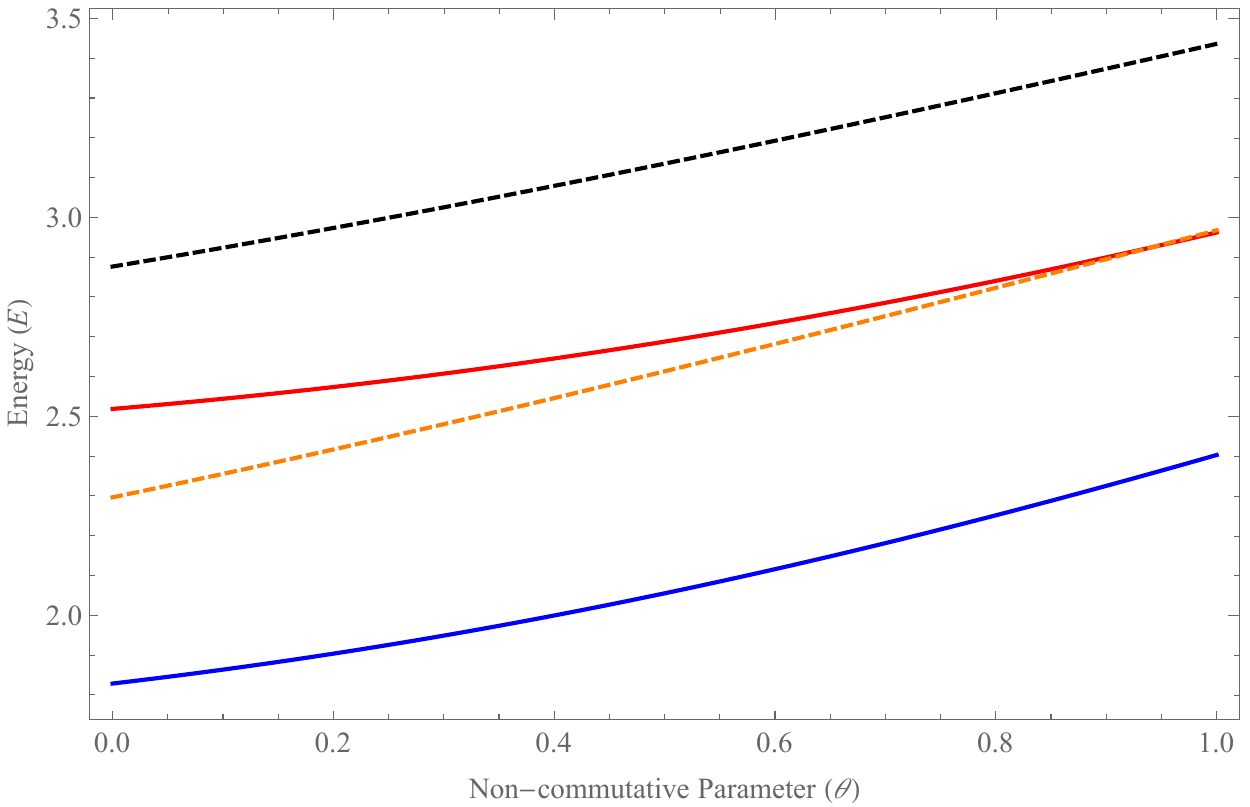}
    \caption{Energy spectrum $E$ as a function of the spatial non-commutative parameter $\theta$. The fixed constants are $\eta=0.1$, $\mu_x=0.5$, and $\mu_y=0.25$ (all evaluated at $\ell=1$). The solid lines represent the system subjected to a magnetic field ($B=1$) for the $(1,1)$ parity state (blue) and the $(-1,-1)$ parity state (red). The dashed lines represent the field-free case ($B=0$) for the $(1,1)$ parity state (orange) and the $(-1,-1)$ parity state (black).}
    \label{fig:theta}
\end{figure}

Finally, Figure \ref{fig:theta} illustrates the energy shift caused by the spatial non-commutativity $\theta$. As $\theta$ grows, the energy of all states consistently increases. Moreover, comparing the magnetized ($B=1$, solid blue and red lines) and field-free ($B=0$, dashed orange and black lines) cases reveals two clear physical effects. First, the magnetized curves are shifted to lower energies compared to the field-free ones, which is consistent with the energy minimum observed at weak magnetic fields in Figure \ref{fig:magnetic}. Second, while the field-free states grow almost linearly with $\theta$, the magnetized states exhibit a pronounced non-linear upward curvature. This confirms that the external magnetic field directly modifies the non-commutative behavior of the system.

\section{Conclusions}

In this work, we obtained the exact analytical solutions for the $(2+1)$-dimensional Klein-Gordon oscillator coupled to a uniform magnetic field, generalized via the Dunkl derivative and extended to a non-commutative phase space. We explicitly calculated the exact energy spectra and eigenfunctions of the Dunkl-Klein-Gordon oscillator in a non-commutative phase space, both with and without a uniform magnetic field. Furthermore, to ensure that our system correctly reduces to the models previously studied in Refs. \cite{Nos2021,Nos2021b}, it was necessary to apply the two transformations in a specific sequence, first introducing the Dunkl derivative, and then applying the non-commutative phase space.

To verify our generalized mathematical framework, we evaluated its limiting reductions. We showed that when the external magnetic field vanishes ($B=0$), our main Eq. (\ref{dkgf}) reduces to the field-free oscillator case given by Eq. (\ref{completaxx}), and the energy spectra of Section 3 reduce to those of Section 2. Furthermore, if the non-commutative parameters are turned off ($\theta=\eta=0$) and $B=0$, our results reduce to the generalized Klein-Gordon equation with the Dunkl derivative \cite{Nos2021}. Additionally, if only $\theta=\eta=0$, our exact solutions recover the commutative Landau levels generalized with the Dunkl derivative \cite{Nos2021b}.

Finally, the discussion of the exact spectra and our graphical analysis show how the phase space geometry and the spatial parity affect the physical system, generalizing previous works \cite{Cai2011,Jian-Hua2008}. In standard non-commutative models, the parameters $\theta$ and $\eta$ usually introduce a global shift in the frequency and the orbital angular momentum. However, in our model, the Dunkl derivative connects these non-commutative scale factors directly with the parity of the system. As demonstrated by Eqs. (\ref{E1CM}) and (\ref{E2CM}), the combination of the non-commutative parameters, the external magnetic field, and the Dunkl reflections creates a parity-dependent modification of the Landau levels. Consequently, scalar bosons with identical quantum numbers but different spatial parities experience different interactions, separating them into distinct energy branches.

\section*{Appendix}

In this appendix, we summarize the results reported in Refs. \cite{Genest2013,Genest2015} regarding the eigenvalues and eigenfunctions of the Dunkl angular momentum operator $\mathcal{J}\equiv i(xD_y-yD_x)$. In polar coordinates, this operator takes the form
\begin{equation}
\mathcal{J}=i(\partial_\phi+\mu_y\cot\phi(1-R_y)-\mu_x\tan\phi(1-R_x)).\label{j}
\end{equation}
A direct calculation shows that the square of this operator is given by the expression
\begin{equation}
\mathcal{J}^2=2B_\phi+2\mu_x\mu_y(1-R_x R_y),\label{jcuad}
\end{equation}
where the angular operator $B_\phi$ is explicitly defined as
\begin{equation}
B_\phi\equiv-\frac{1}{2}\frac{\partial^2}{\partial \phi^2}+\left(\mu_x\tan{\phi}-\mu_y\cot{\phi}\right)\frac{\partial}{\partial \phi}
+\frac{\mu_x (1-R_x)}{2\cos^2{\phi}}+\frac{\mu_y (1-R_y)}{2\sin^2{\phi}}.
\end{equation}

Since the operator $\mathcal{J}$ commutes with $R_xR_y$, and the Hamiltonian $H_O$ given in Eq. (\ref{hcero}) also satisfies $[R_x R_y,H_O]=0$, we can find common eigenvectors that satisfy the eigenvalue equation
\begin{equation}
\mathcal{J}F_\epsilon=\lambda_\epsilon F_\epsilon,\label{angular}
\end{equation}
where $\epsilon\equiv s_x s_y=\pm 1$, with $s_x$ and $s_y$ denoting the eigenvalues corresponding to the reflection operators $R_x$ and $R_y$, respectively. This classification separates the eigensystem into two distinct parity cases.

\begin{itemize}

\item When $s_x=s_y$ (hence $\epsilon=1$), the solutions to Eq. (\ref{angular}) are given by
\begin{eqnarray}
F_+=\Phi_\ell^{++}(\phi)\pm i\Phi_\ell^{--}(\phi),\hspace{.5ex}\\
\lambda_+=\pm2\sqrt{\ell(\ell+\mu_x+\mu_y)},
\end{eqnarray}
where $\ell\in {\mathbb{N}}$. The explicit forms of the functions $\Phi_\ell^{++}$ and $\Phi_\ell^{--}$ are
\begin{eqnarray}
\Phi_\ell^{++}(x)=\sqrt{\frac{(2\ell+\mu_x+\mu_y)\Gamma{(\ell+\mu_x+\mu_y)\ell !}}{2\Gamma{(\ell+\mu_x+1/2)}\Gamma{(\ell+\mu_y+1/2)}}}P_\ell^{(\mu_x-1/2,\mu_y-1/2)}(x),\hspace{18ex}\label{masmas}\\
\Phi_\ell^{--}(x)=\sqrt{\frac{(2\ell+\mu_x+\mu_y)\Gamma{(\ell+\mu_x+\mu_y+1)(\ell-1)!}}{2\Gamma{(\ell+\mu_x+1/2)}\Gamma{(\ell+\mu_y+1/2)}}}\sin\phi\cos\phi P_{\ell-1}^{(\mu_x+1/2,\mu_y+1/2)}(x),\label{menmen}
\end{eqnarray}
where $P_\ell^{(\alpha,\beta)}(x)$ denotes the standard Jacobi polynomials evaluated at $x=-\cos2\phi$. Note that $P_{-1}^{(\alpha,\beta)}(x)$ vanishes by definition, which implies $\Phi_0^{--}=0$.

\item When $s_x=-s_y$ (meaning $\epsilon=-1$), the corresponding solutions are
\begin{eqnarray}
F_-=\Phi_\ell^{-+}(\phi)\mp i\Phi_\ell^{+-}(\phi),\hspace{3ex}\\
\lambda_-=\pm2\sqrt{(\ell+\mu_x)(\ell+\mu_y)},
\end{eqnarray}
where $\ell\in \{\frac{1}{2},\frac{3}{2},...\}$. The associated eigenfunctions $\Phi_\ell^{-+}$ and $\Phi_\ell^{+-}$ take the form
\begin{eqnarray}
\Phi_\ell^{-+}(x)=\sqrt{\frac{(2\ell+\mu_x+\mu_y)\Gamma{(\ell+\mu_x+\mu_y+1/2)(\ell-1/2)!}}{2\Gamma{(\ell+\mu_x+1)}\Gamma{(\ell+\mu_y)}}}\cos\phi P_{\ell-1/2}^{(\mu_x+1/2,\mu_y-1/2)}(x),\label{menmas}\\
\Phi_\ell^{+-}(x)=\sqrt{\frac{(2\ell+\mu_x+\mu_y)\Gamma{(\ell+\mu_x+\mu_y+1/2)(\ell-1/2)!}}{2\Gamma{(\ell+\mu_x)}\Gamma{(\ell+\mu_y+1)}}}\sin\phi P_{\ell-1/2}^{(\mu_x-1/2,\mu_y+1/2)}(x).\hspace{.5ex}\label{masmen}
\end{eqnarray}
\end{itemize}

\section*{Acknowledgments}

%The authors sincerely thank the anonymous reviewers for their careful reading of the manuscript and for their constructive comments and suggestions, which have helped improve the clarity and quality of the work.

This work was partially supported by SNII-M\'exico, COFAA-IPN, EDI-IPN, and CGPI-IPN Project Number IND-2026-0579.

\section*{Disclosures}

The authors declare no conflicts of interest.

\section*{Data Availability}

No data were generated or analyzed in the presented research.

\end{document}